\documentclass[11pt]{article}
\usepackage{amsmath}
\usepackage{amssymb}
\usepackage{graphicx}
\usepackage{verbatim}
\usepackage{psfrag}
\usepackage{listings}
\usepackage{amsmath}            % liefert viele Tools f{\"u}r Math-Mode
\usepackage{amssymb}            % liefert viele Tools f{\"u}r Math-Mode
\usepackage{setspace}           % \onehalfspacing but not in footnotes
\usepackage{verbatim}           % comment environment
\usepackage{graphicx}           % zum Einbinden von Grafiken
\usepackage{psfrag}             % zur Ersetzung von Texten in Grafiken
\usepackage{array}              % defines \extrarowheight
\makeatletter \@addtoreset{equation}{section} \makeatother

\makeatletter
\let\old@startsection=\@startsection
\renewcommand{\@startsection}[6]{\old@startsection{#1}{#2}{#3}{#4}{#5}{#6\mathversion{bold}}}
\makeatother

\DeclareMathOperator{\tr}{tr}                               % su(N) trace operator
\newcommand{\be}{\begin{equation}}
\newcommand{\ee}{\end{equation}}
\newcommand{\ba}{\begin{eqnarray}}
\newcommand{\ea}{\end{eqnarray}}

\newcommand{\setN}{\mathbb{N}}                          % natural
\newcommand{\setZ}{\mathbb{Z}}                          % integer
\newcommand{\unit}{\mathbf{1}}                              % unit element

\newcommand{\susyN}{\mathcal{N}}                             % number of susys
\newcommand{\grU}{{\mathrm{U}}}                             % lie group U
\newcommand{\grSU}{{\mathrm{SU}}}                           % lie group SU
\newcommand{\grSO}{{\mathrm{SO}}}                           % lie group SO
\newcommand{\algU}{{\mathfrak{u}}}                          % lie algebra u
\newcommand{\algSU}{{\mathfrak{su}}}                        % lie algebra su
\newcommand{\algSO}{{\mathfrak{so}}}                        % lie algebra so
\newcommand{\modulus}[1]{{| #1 |}}                          % modulus

\newcommand{\comma}{\quad\mbox{,}\quad}                     % ___,___ in equations
\newcommand{\txt}[1]{\quad\mbox{#1}\quad}                   % ___ text ___ in equations
\newcommand{\ti}[1]{{\scriptscriptstyle\mathrm{#1}}} % textindex
\newcommand{\bibtitle}[1]{{\em #1}}
\newcommand{\hepth}[1]{{\tt hep-th/#1}} 
 
\newcommand{\mathph}[1]{{\tt math-ph/#1}}
\newcommand{\condmat}[1]{{\tt cond-mat/#1}}
\newcommand{\appref}[1]{App.~\ref{#1}}                      % App. #
\newcommand{\tabref}[1]{Tab.~\ref{#1}}                      % Tab. #
\newcommand{\figref}[1]{Fig.~\ref{#1}}                      % Fig. #
\newcommand{\lineref}[1]{line~\ref{#1}}                     % line #
\newcommand{\linesref}[1]{lines~\ref{#1:A}--\ref{#1:B}}     % lines #--#
\newcommand{\code}[1]{\texttt{#1}}                          % program code

\newcommand{\Form}{\texttt{Form}}

\newcommand{\atopfrac}[2]{\genfrac{}{}{0pt}{}{#1}{#2}}
\newcommand{\La}{}                                          % perturbative order, redefined in the main text
\newcommand{\Lb}{\Lambda_r}
\newcommand{\Lc}{\Lambda_r^2}
\newcommand{\Ld}{\Lambda_r^3}

\newcommand{\gym}{g_\ti{YM}}                                % coupling constant g_YM
\newcommand{\gen}[1]{\{#1\}}                                      % Planar su(2) generator
\newcommand{\pD}{\mathsf{D}}                                      % Planar dilatation operator
\newcommand{\pI}{\mathsf{I}}                                      % Planar identity
\newcommand{\pP}{\mathsf{P}}                                      % Planar permutation
\newcommand{\pK}{\mathsf{K}}                                      % Planar trace
\newcommand{\Proj}{\mathit{\Pi}}                                     % projector onto subspace with particular free energy value
\newcommand{\ket}[1]{\bigl|#1\bigr>}                        % Ket vector
\newcommand{\bracket}[3]{\bigl<#1|#2|#3\bigr>}              % Bra C Ket
\newcommand{\txtket}[1]{\ket{\mathrm{#1}}}                  % text Let
\newcommand{\Op}[1]{\bigl|{{\cal O}}_{#1}\bigr>}            % Operator |O>
\newcommand{\comm}[2]{[#1,#2]}                              % commutator
\newcommand{\ad}{a^\dag}
\newcommand{\pp}{\phi}
\newcommand{\PP}{\phi^\dag}

\newcommand{\Z}{Z}
\newcommand{\W}{W}
\newcommand{\Zd}{Z^\dag}
\newcommand{\Wd}{W^\dag}

\begin{document}

%%%%%%% titlepage %%%%%%%

\thispagestyle{empty}
\begin{flushright}
{\sc\footnotesize hep-th/0507217} \\
{\sc\footnotesize AEI-2005-127} \\
\end{flushright}
\vspace{.8cm}
\setcounter{footnote}{0}
\begin{center}
{\Large{\bf On the breakdown of perturbative integrability}} \\[2mm]
{\Large{\bf in large N matrix models}} \\[10mm]
{\sc Thomas Klose} \\[10mm]
{\it Max-Planck-Institut f\"ur Gravitationsphysik \\
     Albert-Einstein-Institut \\
     Am M\"uhlenberg 1, D-14476 Potsdam, Germany} \\ [5mm]
{\tt thklose@aei.mpg.de} \\[12mm]
\end{center}

\begin{abstract}
\noindent
We study the perturbative integrability of the planar sector of a massive $\grSU(N)$ matrix quantum mechanical theory with global $\grSO(6)$ invariance and Yang-Mills-like interaction. This model arises as a consistent truncation of maximally supersymmetric Yang-Mills theory on a three-sphere to the lowest modes of the scalar fields. In fact, our studies mimic the current investigations concerning the integrability properties of this gauge theory. Like in the field theory we can prove the planar integrability of the $\grSO(6)$ model at first perturbative order. At higher orders we restrict ourselves to the widely studied $\grSU(2)$ subsector spanned by two complexified scalar fields of the theory. We show that our toy model satisfies all commonly studied integrability requirements such as degeneracies in the spectrum, existence of conserved charges and factorized scattering up to third perturbative order. These are the same qualitative features as the ones found in super Yang-Mills theory, which were enough to conjecture the all-loop integrability of that theory. For the $\grSO(6)$ model, however, we show that these properties are not sufficient to predict higher loop integrability. In fact, we explicitly demonstrate the breakdown of perturbative integrability at fourth order. 
\end{abstract}

\newpage
\tableofcontents
\vspace{10mm}

%%%%%%%%%%%%%%%%%%%%%%%%%%%%%%%%%%%%%%%%%%%%%%%%%%%%%%%%%%%%%%%%%%%%%%%%%%%%%%%%%%%%%%%%%%%%%%%%%%%%%%%%%%%%%%%%%%%%%%%%%%%%%%%%%%%%
%%%%%%%%%%%%%%%%%%%%%%%%%%%%%%%%%%%%%%%%%%%%%%%%%%%%%%%%%%%%%%%%%%%%%%%%%%%%%%%%%%%%%%%%%%%%%%%%%%%%%%%%%%%%%%%%%%%%%%%%%%%%%%%%%%%%
%%%%%%%%%%%%%%%%%%%%%%%%%%%%%%%%%%%%%%%%%%%%%%%%%%%%%%%%%%%%%%%%%%%%%%%%%%%%%%%%%%%%%%%%%%%%%%%%%%%%%%%%%%%%%%%%%%%%%%%%%%%%%%%%%%%%
\section{Introduction}

%%%%%%% integrability is good for AdS/CFT

The existence of integrable structures in large $N$ gauge theories and superstring theory, that were found and investigated intensively in recent years, have led to an enormous progress in the understanding and the verification of the AdS/CFT correspondence~\cite{ads-cft}%
\footnote{For details about integrability in the AdS/CFT correspondence we would like to refer to the reviews \cite{b:phd,integrability-ads-cft-reviews} and the references therein.}%
. Above all there are the Bethe ansatz techniques which represent novel methods for computing the spectra on either side of the correspondence, i.e. the energy spectrum of string states on the AdS-side and the spectrum of conformal dimensions on the CFT-side. Hence, integrability allows for new tests of the AdS/CFT correspondence by making new data available. Moreover, one can even directly compare the integrable structures and the algebraic equations which encode the spectrum, i.e. an explicit computation of the spectrum can actually be evaded. The performed comparisons have predominantly confirmed the AdS/CFT conjecture, although some discrepancies still need to be resolved. \\

%%%%%%% meaning of integrability in SYM

\begin{figure}[t]
\begin{center}
\small $\tr(ZZZ\,\phi_a\,ZZ\,\psi_\alpha\,ZZ\,D_\mu\,Z) \quad \hat= \quad$
\raisebox{-5mm}{\includegraphics*{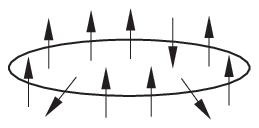}}
$\quad\hat=\quad$
\raisebox{-3mm}{\includegraphics*{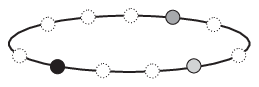}}
\end{center}
\caption[Spin-chain and magnon picture]{\textbf{Spin-chain and magnon picture.} Single-trace SYM operators are in one-to-one correspondence with translationally invariant spin-chain states. Different SYM fields correspond to different spin-alignments or different pseudo-particles called magnons. The highest weight field ($Z$) corresponds to spin up or an empty site.}
\label{fig:spin-chain-picture}
\end{figure}

\noindent In this article we concentrate on the integrable properties discovered on the gauge theory side. Let us therefore briefly recall the findings concerning $\susyN=4$ superconformal $\grSU(N)$ Yang-Mills theory (SYM), which are comprehensively reviewed in~\cite{b:phd}. The statement is that the spectrum of anomalous dimensions of SYM in the 't~Hooft large $N$ limit equals the energy spectrum of an integrable spin-chain system~\cite{mz:bethe-ansatz,bs:sym-spin-chain}. According to this correspondence, single-trace operators of SYM are considered as spin-chain states, cf. \figref{fig:spin-chain-picture}, and the planar part of the SYM dilatation operator $\pD$ becomes the spin-chain Hamiltonian
\be \label{eqn:def-spin-chain-hamiltonian}
  Q_2 = \frac{1}{\lambda} (\pD - \pD_0) \; .
\ee
Herein $\lambda = \gym^2 N$ is the 't~Hooft coupling constant of SYM and $\pD_0$ is the planar dilatation operator of the free theory which measures the bare conformal dimensions. The characteristic feature of the integrability of the spin-chain system is the existence of an infinite tower of higher charges $Q_{r\ge3}$ which commute with each other and with the Hamiltonian%
\footnote{We do not consider the first charge $Q_1$ here. It is usually defined as the spin-chain momentum operator which generates a shift of  all spins by one lattice site. However, as the spin-chain states originate from single-trace SYM operators which are invariant under cyclic permutations of the constituent fields, the momentum operator $Q_1$ acts as a c-number on all considered spin-chain states.}%
:
\be
  \comm{Q_r}{Q_s} = 0 \qquad \forall r,s = 2,3,\ldots \; .
\ee

\begin{figure}[t]
\begin{center}
\psfrag{INSIDE}[c][c]{$\txtket{in}$}
\psfrag{OUTSIDE}[c][c]{$\txtket{out}$'s}
\includegraphics*{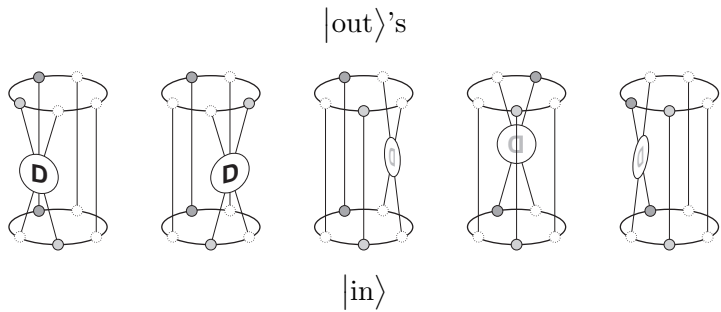}
\end{center}
\caption[Action of dilatation operator]{\textbf{Action of dilatation operator.} This graphics shows the application of a sample dilatation operator of range two to some particular $\txtket{in}$-state. The result is a sum of $\txtket{out}$-states which are obtained by connecting the dilatation operator to all pairs of adjacent fields. In this example the dilatation operator is a mere permutation, and every local application yields just one contribution to the out-state. Generically, the dilatation operator will produce a sum of terms from each local application.}
\label{fig:dilop-action}
\end{figure}

The dilatation operator is not known exactly but must be determined in perturbation theory in the coupling constant $\lambda$~\cite{bks:dilop}. It can be extracted from the logarithmically divergent part of the two-point functions which themselves are computed perturbatively. The application of the dilatation operator to a single-trace SYM operator, alias spin-chain state, corresponds to the attachment of some effective vertex. In the planar limit, this effective vertex is to be connected in all possible ways to a certain number of adjacent fields, cf. \figref{fig:dilop-action}.
The number of involved fields is called the range of the planar dilatation operator. It grows with the perturbative order. At first order in $\lambda$ the dilatation operator has range two, i.e. the associated spin-chain Hamiltonian consists only of nearest neighbor interactions. The higher perturbative contributions to the dilatation operator then cause long-range deformations of the spin-chain Hamiltonian. Notice that due to the extra factor of $\frac{1}{\lambda}$ in \eqref{eqn:def-spin-chain-hamiltonian}, the loop counting in SYM and the spin-chain picture is shifted by one unit, in particular the one-loop dilatation operator corresponds to the zeroth order (i.e. $\lambda = 0$) spin-chain Hamiltonian. 

Likewise to the dilatation operator, all higher charges also receive perturbative correction of growing range. The general expansion is
\be
  Q_r = \sum_{k=0}^\infty \lambda^{k/2} Q_{r,k} \txt{for} r=2,3,\ldots \; ,
\ee
where the range of $Q_{r,k}$ depends essentially linearly on $r$ and $k$. Since the charges are only known up to some perturbative order, we also deal with integrability in a perturbative sense. E.g. when cut off at a certain order in $\lambda$, the charges do not commute exactly with each other but only up to contributions of higher order. Only the lowest order is special in this respect since here $\lambda = 0$ and the commutation of the free charges is exact. \\

%%%%%%% proof of integrability

\noindent Let us also recall how a rigorous proof of integrability would proceed. The reasoning is called \emph{algebraic Bethe ansatz} and is reviewed in~\cite{f:intro-bethe-ansatz}. The main ingredient is the monodromy matrix, or $T$-matrix, which is a spin-chain operator depending on a spectral parameter $u$. The $T$-matrix satisfies a so-called Yang-Baxter equation, which implies that the trace $\tr T$ commutes for different values of the spectral parameter with itself:
\be \label{eqn:monodromy-matrix-commutation}
  \comm{\tr T(u)}{\tr T(u')} = 0 \; . 
\ee
This fact qualifies $\tr T(u)$ as a generating function for an infinite set of commuting charges, which can be obtained by a series expansion of this operator in the spectral parameter.
% It is arbitrary which operator one chooses to define the charges,
% but as $\tr T(u)$ and $\det_q T(u')$ do not commute with each other in general,
% one cannot take the charges from both operators at the same time.

If one of the commuting charges, or a combination of them, coincides with the spin-chain Hamiltonian, then the integrability of the corresponding system is proven. Beyond providing the charges, the algebraic Bethe ansatz also yields algebraic equations, the Bethe equations, which determine the eigenvalues of the charges---in particular the eigenvalues of $Q_2$, i.e. the energy spectrum. \\

%%%%%%% what you do if you cannot prove it

\noindent In practice, however, performing this kind of proof might be rather involved. The difficulty lies in finding the appropriate $T$-matrix and the way to extract the spin-chain Hamiltonian from it. There is no canonical prescription for doing that and even the Yang-Baxter equation is not unique. Therefore, one sometimes tries to find higher commuting charges by hand. In fact, there are arguments that the existence of the first higher charge $Q_3$ already implies integrability~\cite{gm:q3-implies-integrability}. And yet before searching for commuting charges, one should have a look at the spectrum. A good indication for the existence of these charges is the presence of degeneracies in the spectrum which are not due to some obvious symmetries of the system~\cite{bks:dilop}. 

Alternatively, one can also utilize a \emph{coordinate Bethe ansatz}, which is nicely reviewed in~\cite{km:bethe-ansatz}. It is an ansatz for the wave function describing an eigenstate of the spin-chain Hamiltonian. The physical picture behind this ansatz is the propagation of the magnons (see \figref{fig:spin-chain-picture}) along the spin-chain and their scattering. The scattering is encoded in an ``S-matrix'', which is a scalar function of the momenta of the magnons involved in a scattering process. The system is integrable if the multi-particle S-matrix factorizes into a product of two-particle S-matrices. This point of view was stressed and discussed in~\cite{s:paba}. \\

%%%%%%% status in SYM

\noindent The current status of evidence for integrability in SYM is the following. At first order in perturbation theory it has been proven by means of an algebraic Bethe ansatz that the planar SYM dilatation operator~\cite{b:dilop-one-loop} is an $\algSU(4|2,2)$ integrable super spin-chain with nearest neighbor interactions~\cite{bs:sym-spin-chain}. At higher loop orders a similar proof is not known yet. However, the degeneracies in the spectrum, a number of commuting charges and the Bethe equations have been found in different subsectors of the theory: In the $\algSU(3|2)$ subsector, the dilatation operator has been constructed up to third order on the basis of the symmetry algebra, some basic facts about Feynman diagrams and the BMN scaling behavior of the eigenvalues~\cite{b:su32}. BMN scaling is a property predicted by the dual string theory and verified for SYM up to three loops in~\cite{ejs:bmn-3loop}. In the $\algSU(2)$ subsector it was furthermore possible to map the planar three-loop dilatation operator to a known long-range integrable spin-chain~\cite{ss:inozemtsev}, the Inozemtsev spin-chain~\cite{i:int-spin-chain}. But if this equivalence between the SYM dilatation operator and the Inozemtsev spin-chain was supposed to hold also at higher loops, then BMN scaling must be broken in the gauge theory~\cite{ss:inozemtsev}. Therefore one has studied the mutual influence and the compatibility of integrability and BMN scaling, and Beisert, Dippel and Staudacher (BDS) succeeded in writing down an extension of the $\algSU(2)$ spin-chain Hamiltonian up to five-loop order which obeys BMN scaling and commutes with higher charges~\cite{bds:long-range}. Moreover, BDS proposed asymptotic Bethe equations and eigenvalue formulas for all higher charges to all-loop order%
\footnote{\label{footnote:wrapping} ``Asymptotic'' refers to the fact that these equations and formulas are valid only for states longer than a certain threshold which increases with the considered order in the coupling constant. The finite dimensional subsector of shorter states is currently still inaccessible because of the so-called wrapping problem~\cite{bks:dilop,bds:long-range}, which got its name from the fact that it is due to interactions that wrap entirely around the states.}%
. Up to five loops these equations reproduce the spectrum of the explicit Hamiltonian, and beyond five loops they define a novel spin-chain system. As the system is specified in terms of Bethe equations, the BDS spin-chain is in a sense integrable by definition. However, a proof that the charges and the Bethe equations can be derived from a $T$-matrix is still missing, let alone the proof that the BDS spin-chain indeed describes planar SYM in the $\algSU(2)$ subsector. \\

%%%%%%% PWMT as toy model for SYM, status in PWMT

\noindent In a parallel development, the planar integrability of plane-wave matrix theory (PWMT)---a matrix model description of M-theory on a plane-wave~\cite{bmn:bmn-correspondence}---was found~\cite{kp:int-of-pwmt}. PWMT is very closely related to SYM~\cite{kkp:pwmt-from-sym} and concerning integrability it serves as a toy model for the field theory. The energy operator of PWMT, which will be defined as a similarity transformation of the PWMT Hamiltonian below, corresponds to the dilatation operator of SYM. As a matter of fact, the planar energy operator coincides with the planar dilatation operator up to three-loop level in the $\algSU(3|2)$ subsector after appropriate identification of the parameters~\cite{b:su32,kp:int-of-pwmt}. Hence all results concerning integrability that were obtained in SYM up to third order immediately carry over to PWMT. But as PWMT is a quantum mechanical theory, explicit calculations could be pushed to fourth perturbative order in the $\algSU(2)$ subsector of that theory~\cite{fkp:planar4loop}. These computations showed that all integrability criteria such as a degenerate spectrum, existence of conserved charges and factorized scattering persist within PWMT to four-loop order, whereas BMN scaling gets violated. It also turned out that the spin-chain associated to planar PWMT is neither equivalent to the BDS-spin-chain nor to the Inozemtsev spin-chain. Hence, if PWMT should be exactly integrable in the planar sector, it would define a further long-range integrable spin-chain  system. \\

%%%%%%% Current important questions, and this article

\noindent In summary, there are two large $N$ matrix theories, planar SYM and planar PWMT, which display integrable features to relatively high perturbative order. Moreover, there is the beautiful all-loop conjecture of BDS. But despite the tempting simplicity of their formulas, it is strictly speaking absolutely unclear why the integrability should extend to all orders or even exist non-perturbatively. In fact, there is a number of important open questions: What are the essential properties a matrix theory must possess for being integrable at large $N$? What is the mechanism for the symmetry enhancement in the planar limit? What is the simplest matrix model leading to a long-range integrable spin-chain?

In order to address these questions we study the planar integrability of an $\grSO(6)$ matrix model. We introduce this model in the next chapter and connect it to PWMT and SYM. Then we investigate the model with the same tools as used in the study of those theories. We compute the planar energy operator in the $\algSU(2)$ subsector up to fourth order. Very interestingly, we find that this simple toy model indeed passes all commonly required integrability checks (degenerate spectrum, commuting charges, factorized scattering) up to and including third order---but not beyond. Without premonition the degeneracies in the spectrum are lifted by the fourth order piece of the energy operator and therefore perturbative integrability abruptly vanishes as well. These findings certainly sound a note of caution also for PWMT and SYM. This example shows that exact integrability cannot be taken for granted even if perturbative integrability reaches to high loop orders.

In the following we present our results in detail. We also review the essentials of the applied methods. A full account for the technical details can be found e.g. in~\cite{fkp:planar4loop,k:phd}.

%%%%%%%%%%%%%%%%%%%%%%%%%%%%%%%%%%%%%%%%%%%%%%%%%%%%%%%%%%%%%%%%%%%%%%%%%%%%%%%%%%%%%%%%%%%%%%%%%%%%%%%%%%%%%%%%%%%%%%%%%%%%%%%%%%%%
%%%%%%%%%%%%%%%%%%%%%%%%%%%%%%%%%%%%%%%%%%%%%%%%%%%%%%%%%%%%%%%%%%%%%%%%%%%%%%%%%%%%%%%%%%%%%%%%%%%%%%%%%%%%%%%%%%%%%%%%%%%%%%%%%%%%
%%%%%%%%%%%%%%%%%%%%%%%%%%%%%%%%%%%%%%%%%%%%%%%%%%%%%%%%%%%%%%%%%%%%%%%%%%%%%%%%%%%%%%%%%%%%%%%%%%%%%%%%%%%%%%%%%%%%%%%%%%%%%%%%%%%%
\section{The $\grSO(6)$ matrix model}

The degrees of freedom of the model under consideration are comprised in a time-dependent $\grSO(6)$ vector ($a=1,\ldots,6$) that takes values in the adjoint representation ($m=1,\ldots,N^2-1$) of $\algSU(N)$:
\be
  X_a^m(t) \; .
\ee
It is convenient to introduce the basis elements $(T^m)^r_s$ of $\algSU(N)$ which are traceless, hermitian matrices carrying a fundamental index $s$ and an anti-fundamental index $r$ with $s,r=1,\ldots,N$. They satisfy the relations
\be \label{eqn:su-n-generators}
  \comm{T^m}{T^n} = i f^{mnp} T^p
  \comma
  \tr T^m T^n = \delta^{mn}
  \txt{and}
  (T^m)^r_s (T^m)^t_u = \delta^r_u \delta^t_s - \frac{1}{N} \delta^r_s \delta^t_u \; .  
\ee
We use these $\grSU(N)$ generators to define the matrix model fields as
\be
  X_a = X_a^m T^m \; .
\ee
Now, we wish to study the model given by the action 
\be
  S = \int\!dt\: \tr \Bigl[
  \tfrac{1}{2} D_t X_a D_t X_a
  - \tfrac{1}{2} \left( \tfrac{M}{2} \right)^2 \tr X_a X_a
  + \tfrac{1}{4} \tr \comm{X_a}{X_b} \comm{X_a}{X_b}
  \Bigr] \; ,\\
\ee
where $D_t = \partial_t - i \comm{\omega}{\;\;}$ is the covariant derivative containing the scalar gauge field $\omega = \omega^m T^m$, and $M$ is a real parameter which sets both the mass scale of the excitations and the inverse interaction strength as we will discuss below. As a matter of fact, this model is a consistent truncation of PWMT to its $\grSO(6)$ sector~\cite{k:phd}. By definition, this means that the equations of motion of PWMT are satisfied by the solutions of the equations of motion of the $\grSO(6)$ model when setting all other PWMT fields therein to zero. As furthermore the PWMT itself is a consistent truncation of SYM~\cite{kkp:pwmt-from-sym}, we have that the $\grSO(6)$ model is also connected to the field theory by such a procedure.

\begin{figure}[t]
\begin{center}
\psfrag{SYM}{SYM}
\psfrag{SUFTT}{$\mathcal{V}_F$ of $\algSU(4|2,2)$}
\psfrag{PWMT}{PWMT}
\psfrag{SUFT}{$\mathbf{17}$ of $\algSU(4|2)$}
\psfrag{SUTT}{$\mathbf{5}$ of}
\psfrag{SUTTB}{$\algSU(3|2)$}
\psfrag{SOSA}{$\mathbf{6}$ of}
\psfrag{SOSB}{$\algSO(6)$}
\psfrag{SUT}{$\mathbf{2}$ of}
\psfrag{SUTB}{$\algSU(2)$}
\includegraphics*[scale=.8]{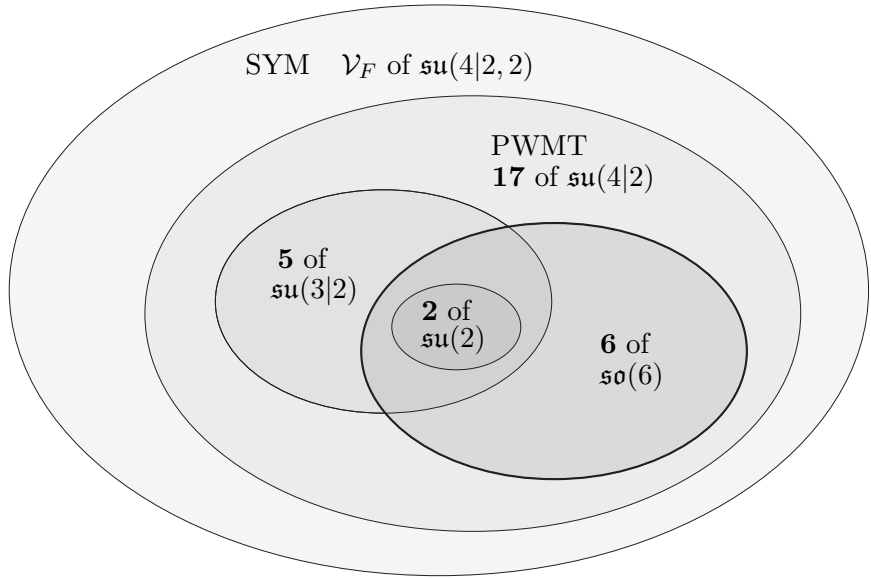}
\end{center}
\caption[Embeddings]{\textbf{Embeddings.} The starting point is the full SYM theory with infinitely many elementary fields in the singleton representation $\mathcal{V}_F$ of $\algSU(4|2,2)$. Seventeen fields can be identified as the degrees of freedom of PWMT. The largest common subsector of SYM and PWMT which is quantum mechanically closed in both theories is spanned by five fields transforming under the fundamental representation of $\algSU(3|2)$. The six fields of the $\grSO(6)$ matrix model represent another common subsector of SYM and PWMT, but one which is not closed in the quantum theories. The $\algSU(2)$ subsector spanned by two complex scalar fields, however, is closed in any of the three models.}
\label{fig:embedding}
\end{figure}

We want to stress that the consistency of a truncation only implies the equivalence of the dynamics of the selected fields at the \emph{classical} level. At the quantum level, however, the fields which are omitted in the reduced theory will generically contribute to the dynamics of the mother theory. Hence, the $\grSO(6)$ model should not be confused with the $\grSO(6)$ subsector of neither PWMT nor SYM. Even within common subsectors, these are really three different quantum theories, which have a connection only at the classical level%
\footnote{It is true that, up to third perturbative order, the planar spectra of PWMT and SYM coincide in the largest common and closed subsector, $\algSU(3|2)$, after appropriately identifying the coupling constants, but this is a highly non-trivial fact, which still needs to be derived from first principles. At any rate, such an equivalence does not exist between the $\grSO(6)$ model and PWMT or SYM.}%
. The relations between the models and their degrees of freedom are depicted in \figref{fig:embedding}.
 
We now fix the gauge $\omega=0$, change to the Hamiltonian formulation, and quantize the model canonically. The Hamiltonian reads 
\be \label{eqn:hamiltonian}
  H = H_0 + V 
\ee
with
\begin{align} \label{eqn:free-ham}
  & H_0 = \tfrac{1}{2} \tr P_a P_a + \tfrac{1}{2} \left( \tfrac{M}{2} \right)^2 \tr X_a X_a
        = \tfrac{M}{2} \tr a_a^\dag a_a \; , \\
  & V   = - \tfrac{1}{4} \tr \comm{X_a}{X_b} \comm{X_a}{X_b} \; .
\end{align}
Here we have already inserted the following mode expansion
\be \label{eqn:mode-expansion}
  P_a = \sqrt{\tfrac{M}{4}} \left( a_a + a_a^\dag \right) \comma
  X_a = \tfrac{i}{\sqrt{M}} \left( a_a - a_a^\dag \right)
\ee
into the free part of the Hamiltonian $H_0$. The modes obey the usual oscillator algebra
\be
  \comm{a_a^m}{a_b^{\dag n}}        = \delta_{ab} \delta^{mn} \comma
  \comm{a_a^m}{a_b^n}               =
  \comm{a_a^{\dag m}}{a_b^{\dag n}} = 0                       \; .
\ee
As a matter of course, the ground state $\ket{0}$ is defined as the unique state in the kernel of all annihilation operators $a_a^m$. The excited states, which are $\grSU(N)$ invariant and relevant in the large $N$ limit, are obtained from this ground state by applying a single-trace of an arbitrary number $L$ of the matrix creation operators $a_a^\dag = a_a^{\dag m} T^m$: 
\be \label{eqn:generic-state}
  \ket{\psi} = \tr ( a_{a_1}^\dag a_{a_2}^\dag \ldots a_{a_L}^\dag ) \ket{0} \; . 
\ee
The number $L$ is called the length of the state. There are no states of length one due to the tracelessness of $\ad_a$. The free energy of \eqref{eqn:generic-state} is $E_0 = \frac{M}{2} L$. The states \eqref{eqn:generic-state} have a spin-chain interpretation as explained in the introduction, cf. \figref{fig:spin-chain-picture}.

Below, we will be particularly interested in an $\algSU(2) \subset \algSO(6)$ subsector of the model. In this sector the elementary oscillators form the following  $\algSU(2)$-doublets 
\be \label{eqn:su2-doublets}
  \PP_i =  \left( { \PP_1 \atop \PP_2 } \right)
        =  \frac{1}{\sqrt{2}} \left( { \ad_1 + i \ad_2 \atop \ad_3 + i \ad_4 } \right)
        =: \left( { \Zd \atop \Wd } \right)
  \comma
  \pp^i =  \left( { \pp^1 \atop \pp^2 } \right)
        =  \frac{1}{\sqrt{2}} \left( { a_1 - i a_2 \atop a_3 - i a_4 } \right)
        =: \left( { \Z \atop \W } \right)
\ee
with commutation relations
\be
  \comm{\pp^{im}}{\pp_j^{\dag n}}       = \delta^i_j \delta^{mn} \comma
  \comm{\pp^{im}}{\pp^{jn}}             =
  \comm{\pp_i^{\dag m}}{\pp_j^{\dag n}} = 0                       \; .
\ee
In the magnon picture, the upper component $\PP_1 \equiv \Zd$ represents an empty side and the lower component $\PP_2 \equiv \Wd$ represents a magnon. In the $\algSU(2)$ subsector there is only one kind of magnon.

%%%%%%%%%%%%%%%%%%%%%%%%%%%%%%%%%%%%%%%%%%%%%%%%%%%%%%%%%%%%%%%%%%%%%%%%%%%%%%%%%%%%%%%%%%%%%%%%%%%%%%%%%%%%%%%%%%%%%%%%%%%%%%%%%%%%
%%%%%%%%%%%%%%%%%%%%%%%%%%%%%%%%%%%%%%%%%%%%%%%%%%%%%%%%%%%%%%%%%%%%%%%%%%%%%%%%%%%%%%%%%%%%%%%%%%%%%%%%%%%%%%%%%%%%%%%%%%%%%%%%%%%%
%%%%%%%%%%%%%%%%%%%%%%%%%%%%%%%%%%%%%%%%%%%%%%%%%%%%%%%%%%%%%%%%%%%%%%%%%%%%%%%%%%%%%%%%%%%%%%%%%%%%%%%%%%%%%%%%%%%%%%%%%%%%%%%%%%%%
\section{Derivation of the associated spin-chain}

As described in the introduction, the planar limit of a matrix theory has a natural interpretation as a spin-chain system. Technically, the spin-chain Hamiltonian is given in essence by the 't~Hooft large $N$ limit of the so-called energy operator. In \cite{fkp:planar4loop} the energy operator
\be \label{eqn:def-enop}
  T = U^{-1} H U
\ee
has been defined as an operator obtained from the Hamiltonian $H$ by a similarity transformation with the property that it does not mix states of different free energy, i.e.
\be \label{eqn:enop-preserves-free-energy}
  \comm{T}{H_0} = 0 \; .
\ee
The primal significance of the energy operator concerns the computation of the quantum mechanical corrections to the free energy spectrum. Due to \eqref{eqn:def-enop} it possesses the same eigenvalues as the full Hamiltonian. However, as it has no overlap between states of different free energy---a consequence of \eqref{eqn:enop-preserves-free-energy}---the energy operator disentangles the mixing problem: only the mixing of states within a degenerate subspace needs to be considered, the influence from states outside is already taken into account in \eqref{eqn:def-enop}.

At this point, however, we want to study the energy operator of the $\grSO(6)$ model with respect to planar integrability. We will use the methods developed for the corresponding investigations in PWMT and SYM. The plan of action now is the following. At first we compute the energy operator $T$ in perturbation theory. Then we define a shifted and rescaled energy operator
\be \label{eqn:redef-enop}
  D := \frac{2}{M} \left( T - \bracket{0}{T}{0} \right)
\ee
in order to have a well-behaved 't~Hooft limit. The planar part $\pD$ of this operator then defines the spin-chain Hamiltonian $Q_2$ similar to \eqref{eqn:def-spin-chain-hamiltonian}. \\

%%%%%%% Computation of enop

\noindent For the computation of the energy operator we use the formulas derived in~\cite{kp:int-of-pwmt,k:phd}. When adopted to the model \eqref{eqn:hamiltonian}, they read
\begin{align} \label{eqn:enop-formula}
T_0 = &\ H_0 \; , \hspace{50mm} \raisebox{0mm}[0mm][0mm]{$\displaystyle T = \sum_{k=0}^\infty T_{2k}$} \\
T_2 = &\ \sum_E \Proj_E V \Proj_E \; , \nonumber \\
T_4 = &\ \sum_E \Proj_E V \Delta_E V \Proj_E \; , \nonumber \\
T_6 = &\ \sum_E \Proj_E \bigl[
                V \Delta_E   V \Delta_E V
              - V \Delta_E^2 V \Proj_E  V
         \bigr] \Proj_E \; , \nonumber \\ 
T_8 = &\ \sum_E \Proj_E \bigl[
                V \Delta_E   V \Delta_E   V \Delta_E V \nonumber \\[-3.5mm]
&\qquad\qquad - V \Delta_E   V \Delta_E^2 V \Proj_E  V
              - V \Delta_E^2 V \Delta_E   V \Proj_E  V
              - V \Delta_E^2 V \Proj_E    V \Delta_E V \nonumber \\
&\qquad\qquad + V \Delta_E^3 V \Proj_E    V \Proj_E  V 
         \bigr] \Proj_E \; , \nonumber
\end{align}
where $\Proj_E$ is a projector onto the subspace of states with free energy $E$, and $\Delta_E$ is the ``propagator'' defined by
\be \label{eqn:def-propagator}
  \Delta_E = \sum_{F\not=E}\frac{\Proj_F}{E - F} \; .
\ee
The sums are taken over all free energies which are realized in this model, i.e. $E \in \frac{M}{2} (\setN\setminus\{1\})$. The piece $T_{2k}$ is called the $k$-th loop contribution. All half-loop contributions are zero in this model, but we maintain the way of indexing the parts of $T$ as in PWMT and SYM for a better comparison. Note that \eqref{eqn:enop-preserves-free-energy} does not uniquely specify the energy operator. In \cite{kp:int-of-pwmt} this ambiguity was fixed by demanding to have the least number of terms. This was an essential requirement to conduct the highly involved computations on current computers but came at the price of the non-hermiticity of the energy operator $T$. However, by a further similarity transformation on top of \eqref{eqn:def-enop} we will change to a hermitian energy operator later. The largest computational effort consists in normal ordering the expressions~\eqref{eqn:enop-formula}. Up to one-loop the result is given by
\begin{align}
T_0 & = \frac{M}{2} \tr \ad_a a_a \; , \\
T_2 & = \frac{1}{M^2} \Bigl[
        15(N^3-N)
        + 10N \tr \ad_a a_a
        + \tfrac{1}{2} : \tr \comm{\ad_a}{a_a} \comm{\ad_b}{a_b} : \nonumber \\
    & \qquad \qquad
        - \tr \comm{\ad_a}{\ad_b} \comm{a_a}{a_b}
        - \tfrac{1}{2} : \tr \comm{\ad_a}{a_b} \comm{\ad_a}{a_b} :
        \Bigr] \; . \label{eqn:enop-one-loop}
\end{align}
We refrain from printing the full, non-planar higher loop contributions as they are rather lengthy. In the $\algSU(2)$ subsector, however, we will give the planar part of the energy operator up to fourth perturbative order, below. \\

%%%%%%% Planar limit

\noindent The next step is to take the planar limit. We will briefly review how this is done at the operatorial level and explain the necessity for the redefinition \eqref{eqn:redef-enop}. As a basic principle, the 't~Hooft limit~\cite{h:large-n} consists of sending the rank of the gauge group to infinity and the coupling constant to zero with the product of these two quantities kept fixed in such a way that physical quantities are neither divergent nor trivial.

The coupling constant of the $\grSO(6)$ model is given by 
\be
  G^2 := \frac{2}{M^3} \; ,
\ee
where the factor of two has been inserted for convenience (and in analogy to PWMT). This can easily be seen from expressing $H_0$ and $V$ in terms of oscillators, see \eqref{eqn:free-ham}--\eqref{eqn:mode-expansion}. It follows that
\be
  H_0 \sim M                    \comma
  V   \sim \frac{1}{M^2}        \comma
  \Proj_E \sim 1                \comma
  \Delta_E \sim \frac{1}{M}     \; .
\ee
Now, either from
\be \label{eqn:find-coupling-constant}
  \frac{V}{H_0} \sim \frac{1}{M^3}
  \quad\txt{or}\quad
  T_{2k} \sim \frac{1}{M^{3k-1}} \sim  \frac{M}{2} \left( \frac{1}{M^3} \right)^k
\ee
we infer that the effective coupling is $\sim \frac{1}{M^3}$. This observation justifies the perturbative treatment of the model for large $M$, and the 't~Hooft limit will also involve $M\to\infty$. In fact, we will have
\be \label{eqn:thooft-like-limit}
  N,M\to\infty \txt{with} \Lambda := G^2 N = \tfrac{2N}{M^3} = \mbox{fixed} \; .
\ee
Note that the parameter $M$ does not only determine the coupling constant but at the same time also represents the energy scale as the global factor of $\frac{M}{2}$ in \eqref{eqn:find-coupling-constant} shows. In order to have a finite limit \eqref{eqn:thooft-like-limit}, we need to rescale the energy operator. This explains the overall factor in \eqref{eqn:redef-enop}. Hence, $D$ measures the energies in units of an elementary excitation.

Beyond that, we also need to analyze the dependence on $N$. All factors of $N$ originate from closed fundamental $\algSU(N)$ index loops, $\tr \unit = \delta^r_r = N$. Apart from the explicit factors in \eqref{eqn:enop-one-loop}, there will arise further powers of $N$ when the operator is applied to a state. Hence, for the purpose of counting its order in $N$, we insert the operator between two states of sufficient length (to avoid wrapping, cf. footnote \ref{footnote:wrapping} on page \pageref{footnote:wrapping}) and normalize with respect to the case where no operator is inserted. The counting is most conveniently done in double line notation. \tabref{tab:operators} shows all effective one-loop vertices in the $\grSO(6)$ model sandwiched between two states.
\begin{table}[t]
\begin{center}
\begin{tabular}{|c||cc|c||c|c|} \hline
 & \multicolumn{3}{c||}{} & \multicolumn{2}{c|}{} \\[-4mm]
% (a) & (b) & (c) & (d) & (e) & (f) \\
free contractions & \multicolumn{3}{c||}{connected} & \multicolumn{2}{c|}{vacuum bubble} \\[1mm]
for reference     & \multicolumn{2}{c|}{planar}    & non-planar           & planar & non-planar \\ \hline
\rule{20mm}{0mm} & \rule{20mm}{0mm} & \rule{20mm}{0mm} & \rule{20mm}{0mm} & \rule{20mm}{0mm} & \rule{20mm}{0mm} \\
%\raisebox{15mm}{vertex:} &
\includegraphics*[scale=.5]{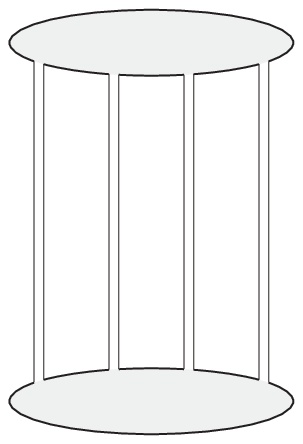}       &
\includegraphics*[scale=.5]{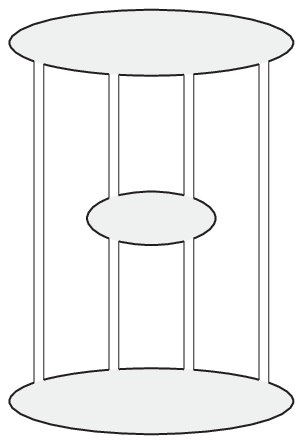}   &
\includegraphics*[scale=.5]{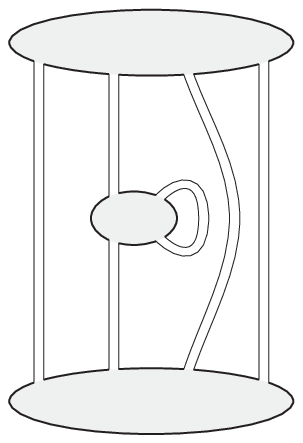}   &
\includegraphics*[scale=.5]{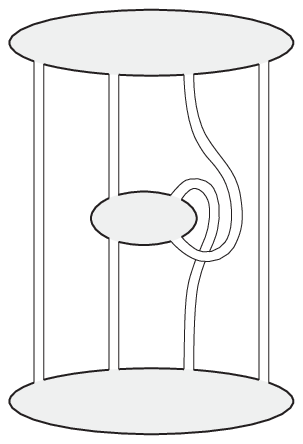} &
\includegraphics*[scale=.5]{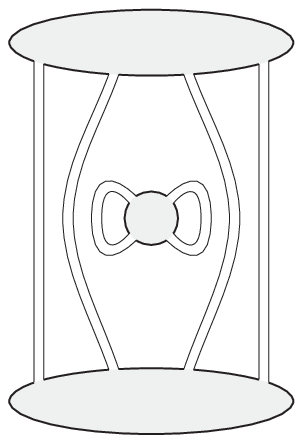}    &
\includegraphics*[scale=.5]{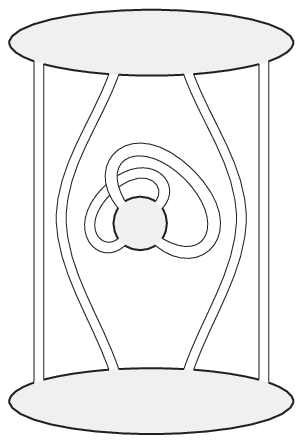} \\[2mm]
%operator: &
$\unit$                  &
$   \tr \ad \ad a \, a$  &
$ N \tr \ad a $          &
$ : \tr \ad a \, \ad a:$ &
$N^3 \cdot \unit$        &
$N   \cdot \unit$ \\[2mm]
%order: &
$\sim 1$   &
$\sim N$   &
$\sim N$   &
$\sim \tfrac{1}{N}$   &
$\sim N^3$ &
$\sim N$ \\[1mm] \hline
\end{tabular}
\end{center}
\caption[One-loop effective vertices]{\textbf{One-loop effective vertices.} If we want to count the powers of $N$ corresponding to a given operator, we insert it into the two-point function and count the additional closed $\algSU(N)$ loops as compared to two-point function without operator insertion.}
\label{tab:operators}
\end{table}
The most important thing to note is the fact that the non-planar graphs are suppressed by factors of $\frac{1}{N^2}$ compared to the planar ones and hence can be eliminated by taking $N$ large. This outcome is nothing than a specific example of the general result of 't~Hooft~\cite{h:large-n}. But one can also observe that the graphs with vacuum bubbles possess an additional factor of $N^2$ with respect to the connected graphs. These contributions can be isolated by computing $\bracket{0}{T}{0}$, and they must be subtracted as in \eqref{eqn:redef-enop} in order to have a finite limit \eqref{eqn:thooft-like-limit}. Physically this corresponds to measuring the energy shifts less the shift of the ground state energy. In fact, this is a reasonable quantity to compute as one can only measure excitations above the ground state. In PWMT, the ground state is protected by supersymmetry and we have exactly $\bracket{0}{T}{0} = 0$.

Now, we can extract the planar part $\pD$ of the redefined energy operator $D$ and, still working at one-loop, we find
\begin{align}
\pD_0 & = \tr \ad_a a_a \; , \hspace{50mm} \raisebox{0mm}[0mm]{$\displaystyle \pD = \sum_{k=0}^\infty \Lambda^k \pD_{2k}$} \\
\pD_2 & = 10 \tr \ad_a a_a
          + \tfrac{1}{N} \tr \ad_a \ad_b a_b a_a
          - \tfrac{2}{N} \tr \ad_a \ad_b a_a a_b
          + \tfrac{1}{N} \tr \ad_a \ad_a a_b a_b \; . \label{eqn:planar-enop-one-loop}
\end{align}
This operator is to be applied to a single-trace state \eqref{eqn:generic-state} in a planar fashion, i.e. the two annihilation operators in the traces of length four have to act onto adjacent creation operators of the state. This application will produce a factor of $N$ which cancels the one in \eqref{eqn:planar-enop-one-loop}. The planar action of $\pD$ can actually be described much simpler by means of the identity operator~$\pI$, the permutation operator~$\pP$ and the trace operator~$\pK$. They are defined to act as
\be
  \pI \, \ad_a \ad_b = \ad_a \ad_b  \comma 
  \pP \, \ad_a \ad_b = \ad_b \ad_a  \comma 
  \pK \, \ad_a \ad_b = \ad_c \ad_c \delta_{ab} \; . 
\ee    
We find
\be
  \pD_0 = \sum_{i=1}^L \; \pI_{i,i+1} = L  \comma
  \pD_2 = \sum_{i=1}^L \bigl[ 11 \, \pI_{i,i+1} - 2 \, \pP_{i,i+1} + \pK_{i,i+1} \bigr] \; .
\ee
The summation runs over the length of the state where $\pD$ is applied to. In indices $()_{i,j}$ mean that the corresponding operator acts onto the $i$-th and $j$-th oscillator. The position $L+1$ is identified with the first position.
 
In this form, the experienced reader will immediately recognize $\pD_2$ as an integrable spin-chain Hamiltonian~\cite{r:integrability}. The integrability hinges on the ratio of the coefficients of $\pP$ and $\pK$. If the spins transform in the vector representation of $\grSO(n)$, integrability requires this ratio to be $-(\tfrac{n}{2}-1)$. For the case at hand, where $n=6$, this condition is satisfied. The integrability of this model is hence completely established in terms of an algebraic Bethe ansatz. The details can be found e.g. in the review~\cite{f:intro-bethe-ansatz} and also in the original paper~\cite{mz:bethe-ansatz} wherein integrability in SYM was discovered. \\

%%%%%%% higher orders

\noindent Let us now proceed to higher loop orders. Their contributions are considered as long-range deformations of the one-loop piece. The full spin-chain Hamiltonian is ``preliminarily'' defined as
\be \label{eqn:def-spin-chain-hamiltonian-recalled}
  Q^\ti{prel}_2 = \frac{1}{\Lambda} (\pD - \pD_0) \; .
\ee
We have called this operator preliminary as we are going to make some minor but convenient redefinitions.

From now on, we will concentrate on the $\algSU(2)$ subsector which is generated by the fields $\PP_i$ defined in \eqref{eqn:su2-doublets}. This entails  a huge simplification, as the trace operator $K$ annihilates all states in this sector. That is basically because there is no invariant tensor $\delta_{ij}$ in $\algSU(2)$. In the evaluation of the expressions \eqref{eqn:enop-formula} we may now discard all terms which contain annihilation operators outside the $\algSU(2)$ subsector. By virtue of being a closed subsector, there are consequently also no terms with creation operator outside this sector. The computation is most straightforward but very cumbersome due to the plethora of terms generated in the process of normal ordering. We have used a \Form{} \cite{v:form} program which does the job for us in roughly 416 hours on a ordinary 2 GHz computer. The program code is printed in \appref{app:code}.

In order to write the result in a compact form, we adopt the frequently used notation for multi-permutation operators
\be \label{eqn:permutations}
  \gen{n_1,n_2,\ldots,n_l} := \sum_{i=1}^L \pP_{i+n_1,i+n_1+1} \pP_{i+n_2,i+n_2+1} \cdots \pP_{i+n_l,i+n_l+1}  \comma
  \gen{}                   := L \; , 
\ee
which was firstly introduced in~\cite{bks:dilop}. There are some obvious relations for these operators
\begin{align}
  \gen{\ldots,n,n,\ldots} & = \gen{\ldots,\ldots} \; ,      \nonumber \\
  \gen{\ldots,n,m,\ldots} & = \gen{\ldots,m,n,\ldots} \; ,  \qquad\qquad \mbox{for $\modulus{n-m} \ge 2$} \\
  \gen{n_1,n_2,\ldots}    & = \gen{n_1+m,n_2+m,\ldots} \; , \nonumber
\end{align}
and another one which is a specialty of the $\algSU(2)$ sector
\be
\begin{split} \label{eqn:permutations-su2-rule}
  & \gen{\ldots,\ldots} + \gen{\ldots,n\pm1,n,\ldots} + \gen{\ldots,n,n\pm1,\ldots} \\
  & - \gen{\ldots,n,\ldots} - \gen{\ldots,n\pm1,\ldots} - \gen{\ldots,n,n\pm1,n,\ldots} = 0 \; .
\end{split}
\ee

Now, we can write down the planar $\algSU(2)$ energy operator of the $\grSO(6)$ model up to fourth perturbative order as
\begin{align} \label{eqn:spin-chain-hamiltonian}
  Q_{2,0} & =   2\gen{}
              - 2\gen{0} \; , \hspace{50mm} \raisebox{0mm}[0mm][0mm]{$\displaystyle Q_2 = \sum_{k=0}^\infty \Lambda^k Q_{2,2k}$} \\
  Q_{2,2} & = - 45\gen{}
              + 49\gen{0}
              - 2(\gen{0,1}+\gen{1,0}) \; , \nonumber \\
  Q_{2,4} & = \tfrac{6313}{4}\gen{}
              - \tfrac{7225}{4}\gen{0}
              + 116(\gen{0,1} + \gen{1,0}) \nonumber \\
& \quad       + 4\gen{0,2}
              - 4(\gen{0,1,2} + \gen{2,1,0}) \; , \nonumber \\
  Q_{2,6} & = -\tfrac{1580065}{24}\gen{}
              + \tfrac{233347}{3}\gen{0}
              - \tfrac{147563}{24}( \gen{0,1} + \gen{1,0} )
              - \tfrac{6089}{16}\gen{0,2} \nonumber \\
& \quad       + \tfrac{5993}{16} ( \gen{0,1,2} + \gen{2,1,0} )
              - \tfrac{87}{16}( \gen{0,2,1} + 220 \gen{1,0,2} )
              - \tfrac{49}{16} \gen{1,0,2,1} \nonumber \\
& \quad       - 4 \gen{0,3}
              + 4 ( \gen{0,1,3} + \gen{0,3,2} )
              + 4 ( \gen{1,0,3} + \gen{0,2,3} )  \nonumber \\
& \quad       - 10 ( \gen{0,1,2,3} + \gen{3,2,1,0} )
              + \tfrac{41}{16} ( \gen{0,1,3,2} + \gen{2,1,0,3} ) \nonumber \\
& \quad       - \tfrac{9}{32}( \gen{0,2,1,3} + \gen{1,0,3,2} + \gen{1,0,2,3} + \gen{0,3,2,1} ) \; . \nonumber
\end{align}
It follows from $Q^\ti{prel}_2$ through
\be \label{eqn:redef-spin-chain-hamiltonian}
  Q_2 = W^{-1} Q^\ti{prel}_2 W - Q_2^\ti{L} \; .
\ee
The similarity transformation by means of the operator
\be
  W(\Lambda) = e^{\Lambda A_1} e^{\Lambda^2 A_2} e^{\Lambda^3 A_3} \; , \\
\ee
where $A_1 = \frac{7}{8} \gen{0}$, $A_2 = \frac{3}{8} \left( \gen{0,1} + \gen{0,1} \right)$, and $A_3 = \frac{209}{16} \left( \gen{0,1} + \gen{0,1} \right)$, corresponds to a change of basis that makes the spin-chain Hamiltonian hermitian. In this notation hermiticity corresponds to the invariance under $\gen{n_1,\ldots,n_l} \mapsto \gen{n_l,\ldots,n_1}$. Furthermore we have subtracted a term proportional to the length operator:
\be
  Q^\ti{L}_2 = \left( 9 - \frac{615}{4} \Lambda + \frac{39123}{8} \Lambda^2 - \frac{37226069}{192} \Lambda^3 \right) \cdot \gen{} \; .
\ee
This operator is determined such that the sum of all coefficients in $Q_2$ vanishes separately at any order. As a consequence $Q_2$ annihilates the states $\tr (\Zd)^L \ket{0}$. The reason for this subtraction is to take away the trivial contribution to the energy proportional to the spin-chain length from the contribution that originates purely from the magnons $\Wd$. Of course, the redefinitions \eqref{eqn:redef-spin-chain-hamiltonian} do not harm integrability; whenever $Q_2$ is an integrable spin-chain Hamiltonian then $Q^\ti{prel}_2$ is one as well and vice versa. Hence we will concentrate on $Q_2$ in the following.

%%%%%%%%%%%%%%%%%%%%%%%%%%%%%%%%%%%%%%%%%%%%%%%%%%%%%%%%%%%%%%%%%%%%%%%%%%%%%%%%%%%%%%%%%%%%%%%%%%%%%%%%%%%%%%%%%%%%%%%%%%%%%%%%%%%%
%%%%%%%%%%%%%%%%%%%%%%%%%%%%%%%%%%%%%%%%%%%%%%%%%%%%%%%%%%%%%%%%%%%%%%%%%%%%%%%%%%%%%%%%%%%%%%%%%%%%%%%%%%%%%%%%%%%%%%%%%%%%%%%%%%%%
%%%%%%%%%%%%%%%%%%%%%%%%%%%%%%%%%%%%%%%%%%%%%%%%%%%%%%%%%%%%%%%%%%%%%%%%%%%%%%%%%%%%%%%%%%%%%%%%%%%%%%%%%%%%%%%%%%%%%%%%%%%%%%%%%%%%
\section{Perturbative integrability and its breakdown}

In the previous section we have seen that the planar one-loop energy operator of the $\grSO(6)$ model taken by itself defines an exactly integrable spin-chain system with nearest neighbor interactions. Higher loop integrability is, however, hard to prove because it is not known how to generalize the monodromy matrix and perhaps also the Yang-Baxter equation appropriately, not to mention how to show that the Hamiltonian is among the commuting charges. This is precisely the same situation as currently in PWMT and SYM. In those cases one therefore concentrates on the symptoms of integrability. These are the degeneracies in the spectra due to the existence of higher charges and the factorization of the S-matrix describing the multi-magnon scattering. These properties are widely accepted as strong evidence for higher loop integrability.

In this section we demonstrate on the one hand that the $\grSO(6)$ model is perturbatively integrable in this sense up to and including third order. On the other hand we also show that at four-loop level the situation is completely changed. A charge that commutes perturbatively with the Hamiltonian does not exist any more and the degeneracies are indeed lifted. A two-magnon S-matrix does, of course, still exist but it does not reproduce the energies of multi-magnon states by means of Bethe equations. \\

%%%%%%% Computation of spectrum

\noindent The computation of the spectrum of the spin-chain Hamiltonian~\eqref{eqn:spin-chain-hamiltonian} proceed as follows. First of all we need to find which states are realized and how they are organized in multiplets. An $\algSU(2)$ multiplet can be labeled by the length $L$ and the magnon number $M$ of the highest weight state. All states of one multiplet have the same length but descendent states possess an increased number of magnons. The multiplicities $m$ of a certain multiplet $(L,M)$ in the spectrum is given by the number of linearly independent $M$-magnon states of length $L$ that are highest weight states, i.e. that are annihilated by the raising operator $J_+$. $J_+$ acts on states according to the Leibniz rule and then on each single oscillators as
\be
  J_+ \Wd = \Zd \comma  J_+ \Zd = 0 \; . 
\ee
The multiplets can furthermore be disentangled such that their states possess definite parity~$p$. The parity conjugation operator $P$ acts on the $\grSU(N)$ generators as transposition and multiplication by $-1$, i.e. on single-traces as
\be
  P \, \tr T^{m_1} \cdots T^{m_L} \, P^{-1} = (-1)^L \tr T^{m_L} \cdots T^{m_1} \; ,
\ee   
and the vacuum has positive parity $P \ket{0} = \ket{0}$. Hence, parity conjugation essentially reverses the oscillators in a state
\be
  P \tr \PP_{i_1} \cdots \PP_{i_L} \ket{0} = (-1)^L \tr \PP_{i_L} \cdots \PP_{i_1} \ket{0} \; .
\ee
The multiplicities of all $\algSU(2)$ multiplets with $L\le9$ together with their parity are listed in \tabref{tab:su2-sector}.
\setlength{\extrarowheight}{2pt}%
\begin{table}[t]
\begin{center}
\begin{footnotesize}
\begin{tabular}{|c|c|c|c|c|c|c|c|c|c|c|c|c|} \hline
$L$   & $4$ & $5$ & \multicolumn{2}{c|}{$6$} & \multicolumn{2}{c|}{$7$} & \multicolumn{3}{c|}{$8$} & \multicolumn{3}{c|}{$9$} \\ \hline
$M$   & $2$ & $2$ & $2$ & $3$ & $2$ & $3$ & $2$ & $3$ & $4$ & $2$ & $3$ & $4$ \\ \hline
$m^p$ & $1^+$ & $1^-$ & $2^+$ & $1^-$ & $2^-$ & {\textbf{\mathversion{bold} $1^\pm$}} & $3^+$ & {\textbf{\mathversion{bold} $1^\pm$}},$1^-$ & $3^+$ & $3^-$ & {\textbf{\mathversion{bold} $3^\pm$}} & {\textbf{\mathversion{bold} $1^\pm$}},$2^-$ \\ \hline
\end{tabular}
\end{footnotesize}
\end{center}
\caption[States in $\algSU(2)$ sector]{\textbf{\mathversion{bold} States in $\algSU(2)$ sector.} Multiplets are labeled by the length $L$, the magnon number $M$ of the highest weight states and the parity $p$. This table lists the multiplicities $m$ of all irreducible multiplets with $L\le9$. Here $m^\pm$ denotes $m$ pairs of multiplets whose partners have opposite parity. These pairs, printed in bold face, have degenerate energies if the associated spin-chain system is integrable.}
\label{tab:su2-sector}
\end{table}
\setlength{\extrarowheight}{0pt}
Of particular importance for integrability are the parity pairs, denoted by $m^\pm$ in the table. The states of a pair, $\ket{+}$ and $\ket{-}$, are related by a charge $Q_3$ which is parity odd, $P Q_3 P^{-1} = - Q_3$, and which commutes with the Hamiltonian, $\comm{Q_2}{Q_3}=0$. This implies that these states have identical energy. In formulas this is
\be
  \ket{+} = Q_3 \ket{-} \comma Q_2\ket{\pm} = \ket{\pm} q_2^\pm  \qquad \Rightarrow \qquad  q_2^+ = q_2^-  \; .
\ee

The systematic occurrence of these degeneracies are a strong indication of the existence of such a commuting charge $Q_3$ and thus also of the presence of integrability. In \tabref{tab:so6-spectrum} we give the energies $q_2$ for all parity pairs up to $L\le9$. The observation is that the partners of a pair have exactly identical energies up to third perturbative order but slightly different ones at fourth order. \\

\renewcommand{\La}{}
\renewcommand{\Lb}{\Lambda}
\renewcommand{\Lc}{\Lambda^2}
\renewcommand{\Ld}{\Lambda^3}
\begin{table}[t]
\begin{center}
\begin{footnotesize}
\begin{tabular}{|c|c|l|l|} \hline
$L$ & $M$ & Spin-chain energy & Bethe momenta \\ \hline
$7$ & $2$ & $q_2 = 4 \La - 80 \Lb + 2595 \Lc - \frac{1231451}{12} \Ld \qquad (-) = \mbox{Bethe}$  &
$\begin{array}{l} p_1 = \tfrac{\pi}{3} \\ p_2 = -p_1 \end{array}$  \\ \cline{3-4}
    &     & $q_2 = 12 \La - 264 \Lb + 9111 \Lc - \frac{1502247}{4} \Ld \qquad (-) = \mbox{Bethe}$ &
$\begin{array}{l} p_1 = \tfrac{2\pi}{3} \\ p_2 = -p_1 \end{array}$ \\ \hline
$7$ & $3$ & $q_2 = 10 \La - 215 \Lb + \frac{29325}{4} \Lc + 
 \begin{cases}
  -\frac{7205509}{24} \Ld                 & (+) \\
  -\frac{7205365}{24} \Ld                 & (-) = \mbox{Bethe}
 \end{cases}$ &
$\begin{array}{l} p_1 = \pm1.16\pm0.93i \\ p_2 = \pm1.16\mp0.93i \\ p_3 = -p_1-p_2 \end{array}$  \\ \hline
$8$ & $3$ & $q_2 = 8 \La - 168 \Lb + 5633 \Lc + 
 \begin{cases}
  -\frac{1367125}{6} \Ld               & (+) = \mbox{Bethe} \\
  -\frac{1367116}{6} \Ld               & (-) \\
 \end{cases}$ &
$\begin{array}{l} p_1 = \pm0.96\pm0.59i \\ p_2 = \pm0.96\mp0.59i \\ p_3 = -p_1-p_2 \end{array}$  \\ \hline
$9$ & $3$ & $q_2 \approx 6.45322 \La - 132.568 \Lb + 4378.58 \Lc + 
 \begin{cases}
  -175265.3 \Ld              & (+) \\
  -175267.0 \Ld              & (-) = \mbox{Bethe} 
 \end{cases}$ &
$\begin{array}{l} p_1 = \mp0.83\mp0.43i \\ p_2 = \mp0.83\pm0.43i \\ p_3 = -p_1-p_2 \end{array}$  \\ \cline{3-4}
    &     & $q_2 \approx 11.0399 \La - 237.742 \Lb + 8122.41 \Lc + 
 \begin{cases}
  -333240.5 \Ld               & (+) \\
  -333230.2 \Ld               & (-) = \mbox{Bethe}
 \end{cases}$ &
$\begin{array}{l} p_1 = \mp1.28\mp1.26i \\ p_2 = \mp1.28\pm1.26i \\ p_3 = -p_1-p_2 \end{array}$  \\ \cline{3-4}
    &     & $q_2 \approx 16.5068 \La - 360.690 \Lb + 12421.3 \Lc + 
 \begin{cases}
  -511941.8 \Ld               & (+) \\
  -511944.4 \Ld               & (-) = \mbox{Bethe}
 \end{cases}$ &
$\begin{array}{l} p_1 = \pm2.98 \\ p_2 = \pm1.15 \\ p_3 = -p_1-p_2 \end{array}$  \\ \hline
$9$ & $4$ & $q_2 = 10 \La - 215 \Lb + \frac{29405}{4} \Lc + 
 \begin{cases}
  -\frac{79730123}{264} \Ld    & (+) \\
  -\frac{79732139}{264} \Ld    & (-) \\
  -\frac{79734083}{264} \Ld    & \mbox{Bethe}
 \end{cases}$ &
$\begin{array}{l} p_1 = \pm2.63 \\ p_2 = \mp0.53\pm0.88i \\ p_3 = \mp1.77 \\ p_4 = -p_1-p_2-p_3 \end{array}$  \\ \hline
\end{tabular}
\end{footnotesize}
\end{center}
\caption[Selected states of $\algSU(2)$ spectrum]{\textbf{\mathversion{bold} Selected states of $\algSU(2)$ spectrum.} We list the eigenvalues as computed by direct application of the Hamiltonian to states and explicit diagonalization. ``(+/--)'' refers to the parity of the states. We observe the degeneracy of the parity pairs up to three-loop and the lift of the degeneracy at four-loop order. We also calculate the eigenvalues by means of the Bethe ansatz. (The values $p_i$ give the $\Lambda^0$-th order of the quasi-momenta.) The result coincides with the elementary calculation only up to three loops.}
\label{tab:so6-spectrum}
\end{table}

%%%%%%% Commuting charge

\noindent We show that this behavior is not just an effect for short lengths $L$ by presenting a local, parity-odd charge $Q_3$ satisfying
\be
  \comm{Q_2}{Q_3} = \mathcal{O}(\Lambda^3)  \quad\txt{(i.e. zero up to fourth order terms)} .
\ee 
It reads
\begin{align}
Q_{3,0} =\ & c_{3,0,1} ( \gen{0,1} - \gen{1,0} ) \; , \hspace{50mm} \raisebox{0mm}[0mm][0mm]{$\displaystyle Q_3 = \sum_{k=0}^2 \Lambda^k Q_{3,2k}$} \\[3mm]
Q_{3,2} =\ & c_{3,2,1} ( \gen{0,1} - \gen{1,0} )
             + 2c_{3,0,1} ( \gen{0,1,2} - \gen{2,1,0} ) \; ,\nonumber \\[3mm]
Q_{3,4} =\ & c_{3,4,1} ( \gen{0,1} - \gen{1,0} )
             + (-57c_{3,0,1}+2c_{3,2,1}-2c_{3,4,2}) ( \gen{0,1,2} - \gen{2,1,0} ) \nonumber \\
           & - 2c_{3,0,1} ( \gen{0,1,3} + \gen{0,2,3} - \gen{0,3,2} - \gen{1,0,3} )
             + c_{3,4,2} ( \gen{0,1,2,3} - \gen{3,2,1,0} ) \nonumber \\
           & + (4c_{3,0,1}-c_{3,4,2}) ( \gen{0,2,1,3} - \gen{1,0,3,2} + \gen{0,1,3,2} - \gen{0,3,2,1} \nonumber \\
           & \hspace{80mm} + \gen{1,0,2,3} - \gen{2,1,0,3} ) \; . \nonumber
\end{align}
In this notation parity conjugation acts as $P \gen{n_1,\ldots,n_l} P^{-1} = \gen{-n_1,\ldots,-n_l}$. The constants~$c$ are not fixed by solely demanding the commutation with $Q_2$. And as to be expected from the lift of the degeneracies, there is \emph{no} parity-odd operator $Q_{3,6}$ of (maximal) range six which satisfies
\be
  \comm{Q_{3,6}}{Q_{2,0}} + \comm{Q_{3,4}}{Q_{2,2}} + \comm{Q_{3,2}}{Q_{2,4}} + \comm{Q_{3,0}}{Q_{2,6}} = 0 \; .
\ee \\

%%%%%%% PABA

\noindent Finally, we consider the perturbative asymptotic Bethe ansatz \cite{s:paba} for this model. We deduce the two-magnon S-matrix from the spin-chain Hamiltonian $Q_2$ and compute from it the energies of all states in \tabref{tab:so6-spectrum} by means of Bethe equations. This will show the factorization of the S-matrix up to three-loop level, but it will not reveal any suspicious behavior of the system that might help to anticipate the breakdown of integrability at fourth order.

We derive the S-matrix according to the method explicitly explained in \cite{fkp:planar4loop,k:phd}. One considers all spin-chain fragments (i.e. no $\algSU(N)$ trace) of length $L$ with 2 magnons at positions $l_1$ and $l_2$
\be
 \Op{l_1,l_2} := (\Zd)^{l_1-1} \Wd (\Zd)^{l_2-l_1-1} \Wd (\Zd)^{L-l_2} \ket{0}   \qquad \mbox{for $1 \le l_1 < l_2 \le L-1$} \; .
\ee
These fragments are superposed to energy eigenstates of the Hamiltonian
\be \label{eqn:eigenvalue-ansatz}
  Q_2 \ket{p_1,p_2} = \ket{p_1,p_2} q_2(p_1,p_2)
\ee
according to the perturbative asymptotic Bethe ansatz as
\be \label{eqn:eigenstate-ansatz}
  \ket{p_1,p_2} := \sum_{\textstyle \atopfrac{l_1,l_2=1}{l_1<l_2}}^L a(l_1,l_2,p_1,p_2) \Op{l_1,l_2}
\ee
with
\be
\begin{split}
  a(l_1,l_2,p_1,p_2) =\ & e^{i(p_1 l_1 + p_2 l_2)} f(l_2-l_1,p_1,p_2) \\
                        & + e^{i(p_1 l_2 + p_2 l_1)} f(L-l_2+l_1,p_1,p_2) S(p_2,p_1) \; .
\end{split}
\ee
The exponentials describe the free propagation of the magnons along the spin-chain with quasi-momenta $p_1$ and $p_2$. The functions
\be
  f(l,p_1,p_2) = 1 + \Lambda^{l} f_0(l,p_1,p_2) + \Lambda^{l+1} f_1(l,p_1,p_2) + \Lambda^{l+2} f_2(l,p_1,p_2) + \ldots
\ee
and
\be
  S(p_1,p_2) = S_0(p_1,p_2) + \Lambda S_1(p_1,p_2) + \Lambda^2 S_2(p_1,p_2) + \Lambda^3 S_3(p_1,p_2) + \ldots \; .
\ee
describe the magnon scattering due to the long-range ($f$) and the nearest neighbor interactions ($S$), respectively. Joining the end of the fragments by taking the trace in order to obtain cyclic spin-chains leads to the two-magnon Bethe equations
\be \label{eqn:2-magnon-bethe-equations}
  \exp(iLp_1) = S(p_1,p_2) \comma \exp(iLp_2) = S(p_2,p_1)
\ee
and the total momentum condition
\be
  p_1 + p_2 = 2 \pi \setZ \; .
\ee
In an integrable system where the multi-magnon S-matrix is factorized into products of the two-magnon S-matrix, these equations have straightforward generalizations to the $M$-magnon case
\be \label{eqn:M-magnon-bethe-equations}
  \exp(iLp_k)
  = \prod_{\textstyle \atopfrac{j=1}{j\not=k}}^M S(p_k,p_j)
  \txt{for $k=1,\ldots,M$}
  \comma
  \sum_{i=1}^M p_i = 2 \pi \setZ \; .
\ee

Specializing to the $\grSO(6)$ model, we act with the spin-chain Hamiltonian \eqref{eqn:spin-chain-hamiltonian} onto \eqref{eqn:eigenstate-ansatz} and demanding \eqref{eqn:eigenvalue-ansatz} fixes all functions $q_2(p_1,p_2)$, $f(l,p_1,p_2)$, and $S(p_1,p_2)$ up to $\mathcal{O}(\Lambda^3)$ (fourth perturbative order). The energy is given by the sum $q_2(p_1,p_2) = q_2(p_1) + q_2(p_2)$ of a one-magnon energy
\be
\begin{split}
q_2(p) =\ & 8 \sin^2(\tfrac{p}{2})
            - \Lambda \left[ 148 \sin^2(\tfrac{p}{2}) + 32 \sin^4(\tfrac{p}{2}) \right] \\
          & + \Lambda^2 \left[ 4601 \sin^2(\tfrac{p}{2}) + 1472 \sin^4(\tfrac{p}{2}) + 256 \sin^6(\tfrac{p}{2}) \right] \\
          & - \Lambda^3 \left[ \tfrac{1066301}{6} \sin^2(\tfrac{p}{2}) + \tfrac{197109}{3} \sin^4(\tfrac{p}{2})
                               + 18816 \sin^6(\tfrac{p}{2}) + 2560 \sin^8(\tfrac{p}{2}) \right] \; .
\end{split}
\ee
In the generalization to $M$ magnons, every magnon contributes a corresponding portion to the total energy
\be \label{eqn:eigenvalue-formula}
  q_2 = \sum_{i=1}^M q_2(p_i) \; .
\ee
The first three orders of the S-matrix can be brought into the usual form
\be \label{eqn:so6-s-matrix}
  S(p_1,p_2) = \frac{\varphi(p_1)-\varphi(p_2)+i}{\varphi(p_1)-\varphi(p_2)-i} + \mathcal{O}(\Lambda^3)
\ee
where here the phase function reads
\be
  \varphi(p) = \tfrac{1}{2}\cot(\tfrac{p}{2}) \bigl[ 1
  + 8 \Lambda   \sin^2(\tfrac{p}{2})
- 220 \Lambda^2 \sin^2(\tfrac{p}{2})
-  32 \Lambda^2 \sin^4(\tfrac{p}{2}) \bigr] \; .
\ee
We refrain from printing the fourth order piece as it is not of this or any other obvious compact form. Also the introduction of an exponential factor into \eqref{eqn:so6-s-matrix} (as in PWMT~\cite{fkp:planar4loop} or on the string theory side~\cite{afs:quantum-strings}) did not lead to a meaningful expression. Moreover, we do not print the function $f(l,p_1,p_2)$ as it is not relevant for the computation of the eigenvalues.

Now, we solve the general Bethe equations \eqref{eqn:M-magnon-bethe-equations} with \eqref{eqn:so6-s-matrix} for the cases of \tabref{tab:so6-spectrum}. The lowest order quasi-momenta are given in the table. Then we compute the energy eigenvalues using \eqref{eqn:eigenvalue-formula} and compare them with the results from the direct diagonalization of the Hamiltonian. Up to third order we find exact agreement. At fourth order the Bethe ansatz approach must fail as, on the one hand, the degeneracies between parity pairs are lifted but, on the other hand, the Bethe ansatz cannot distinguish between the two partners of a parity pair---essentially because $q_2(p) = q_2(-p)$. However, we observe that in the considered three-magnon cases, the Bethe solutions coincides with one state of a pair. This is no longer the case for the four-magnon pair where the Bethe solution differs slightly from the true energies of both states. For the two-magnon states, on the other side, the Bethe ansatz still works at fourth order. This in not surprising as the Bethe equations \eqref{eqn:2-magnon-bethe-equations} in this sector follow from a rigorous derivation; no factorization had to be assumed.

%%%%%%%%%%%%%%%%%%%%%%%%%%%%%%%%%%%%%%%%%%%%%%%%%%%%%%%%%%%%%%%%%%%%%%%%%%%%%%%%%%%%%%%%%%%%%%%%%%%%%%%%%%%%%%%%%%%%%%%%%%%%%%%%%%%%
%%%%%%%%%%%%%%%%%%%%%%%%%%%%%%%%%%%%%%%%%%%%%%%%%%%%%%%%%%%%%%%%%%%%%%%%%%%%%%%%%%%%%%%%%%%%%%%%%%%%%%%%%%%%%%%%%%%%%%%%%%%%%%%%%%%%
%%%%%%%%%%%%%%%%%%%%%%%%%%%%%%%%%%%%%%%%%%%%%%%%%%%%%%%%%%%%%%%%%%%%%%%%%%%%%%%%%%%%%%%%%%%%%%%%%%%%%%%%%%%%%%%%%%%%%%%%%%%%%%%%%%%%
\section{Conclusions}

% TODO: Integrability in SYM because of AdS/CFT, see super-spin-chain paper.

%%%%%%% What we have done

At the moment the approved data concerning integrability in SYM reach up to third~\cite{b:su32} and in PWMT up to fourth perturbative order~\cite{fkp:planar4loop}. In this article we have posed the question whether these data are sufficient to justify the belief in all-loop integrability. In fact, we have presented a toy model where the answer is negative. The model under consideration is the $\grSO(6)$ matrix model, which is related to both PWMT and SYM by a consistent truncation. This model displays the same integrability properties as its mother theories: exact integrability at one-loop order, and degeneracies in the spectrum, existence of conserved charges as well as factorized scattering up to three-loop order. But for all that, as one proceeds to the next order, all of these properties and thus integrability abruptly cease to exist. 

%%%%%%% What it implies

The moral of this article is the simple warning, that despite the presence of perturbative integrability at low orders, one cannot take higher or all-loop integrability for granted. Clearly, also before this example it was evident that integrability may break down at any arbitrary order; just take any long-range integrable spin-chain, e.g. the Inozemtsev spin-chain, and add a deformation that breaks integrability at some order. The point here is that we did not use an artificially constructed spin-chain but started from a decent looking matrix theory which pretended to be integrable.

%%%%%%% Being realistic and honest: 

Admittedly, the $\grSO(6)$ matrix model has very much less structure than SYM and also PWMT. On the one hand, the additional symmetries of the latter theories might be just what is needed for exact integrability. From the three-loop investigations concerning the $\algSU(3|2)$ sector of SYM~\cite{b:su32}, we know that the symmetry (together with BMN scaling) fixes the dilatation operator strongly enough to imply planar integrability. However, the higher the perturbative order the weaker are the restrictions from symmetry. So, on the other hand, the additional symmetries may just shift the breakdown of integrability to a higher level. Therefore it is essential to know how much freedom the symmetries ultimately leave and what the basic cause for integrability really is. 

%%%%%%% Future directions

It does not lead to any principle improvement if we could push the perturbative results by one or two orders (unless we find the breakdown of integrability) and we eventually need to verify the planar integrability non-perturbatively. In the field theory, however, this is a rather formidable aim since the dilatation operator is not known exactly. Therefore we propose to intensify the study of quantum mechanical matrix theories where the full Hamiltonian is given. A particularly interesting theory would be PWMT but other appropriate toy models can be constructed as well. In possession of the full Hamiltonian, one might be able to go without invoking perturbation theory and one can try to determine charges that commute exactly. An important prerequisite for such a program, however, would be to implement the planar limit at the level of the Hamiltonian. This is a highly non-trivial task but it is of central and essential significance for proving integrability in large $N$ matrix theories.

%%%%%%%%%%%%%%%%%%%%%%%%%%%%%%%%%%%%%%%%%%%%%%%%%%%%%%%%%%%%%%%%%%%%%%%%%%%%%%%%%%%%%%%%%%%%%%%%%%%%%%%%%%%%%%%%%%%%%%%%%%%%%%%%%%%%
%%%%%%%%%%%%%%%%%%%%%%%%%%%%%%%%%%%%%%%%%%%%%%%%%%%%%%%%%%%%%%%%%%%%%%%%%%%%%%%%%%%%%%%%%%%%%%%%%%%%%%%%%%%%%%%%%%%%%%%%%%%%%%%%%%%%
%%%%%%%%%%%%%%%%%%%%%%%%%%%%%%%%%%%%%%%%%%%%%%%%%%%%%%%%%%%%%%%%%%%%%%%%%%%%%%%%%%%%%%%%%%%%%%%%%%%%%%%%%%%%%%%%%%%%%%%%%%%%%%%%%%%%
\section*{Acknowledgments}

It is my pleasure to thank Abishek Agarwal, Niklas Beisert, Jan Plefka, and Matthias Staudacher for interesting and stimulating discussions, and for their valuable comments on the manuscript.

%%%%%%%%%%%%%%%%%%%%%%%%%%%%%%%%%%%%%%%%%%%%%%%%%%%%%%%%%%%%%%%%%%%%%%%%%%%%%%%%%%%%%%%%%%%%%%%%%%%%%%%%%%%%%%%%%%%%%%%%%%%%%%%%%%%%
%%%%%%%%%%%%%%%%%%%%%%%%%%%%%%%%%%%%%%%%%%%%%%%%%%%%%%%%%%%%%%%%%%%%%%%%%%%%%%%%%%%%%%%%%%%%%%%%%%%%%%%%%%%%%%%%%%%%%%%%%%%%%%%%%%%%
%%%%%%%%%%%%%%%%%%%%%%%%%%%%%%%%%%%%%%%%%%%%%%%%%%%%%%%%%%%%%%%%%%%%%%%%%%%%%%%%%%%%%%%%%%%%%%%%%%%%%%%%%%%%%%%%%%%%%%%%%%%%%%%%%%%%
\appendix
\section{Program code} \label{app:code}

We print the computer program which was used to normal order the energy operator $T$ as defined in \eqref{eqn:enop-formula}. It is written in \Form{}~\cite{v:form} and the required interpreter can be downloaded from http://www.nikhef.nl/$\tilde{\;\;}$form/.

Starting point is the definition of the energy operator in \linesref{lin:enop}. Here, \code{In} marks the place where the initial state with free energy $E_0$ will be attached, and \code{D(p,e)} stands for the $p$-th power of the propagator $\Delta_{E_0+e}$ \eqref{eqn:def-propagator} if $p\ge1$ and for the projector $\Proj_{E_0+e}$ if $p=-1$. The vertex \code{V} is expressed in terms of the matrix model field $\code{X(a,m)} \leftrightarrow X_a^m$ in \linesref{lin:substitute-vertex} and subsequently in terms of modes $\code{A(a,m,+1)} \leftrightarrow a_a^{\dag m}$, $\code{A(a,m,-1)} \leftrightarrow a_a^m$. The mass parameter $M$ is set to unit during this computation. Now, we can evaluate the propagators and projectors (\linesref{lin:eval-prop-proj}) and perform the normal ordering of the expression (\linesref{lin:nod}). In \lineref{lin:is-nod} we turn the oscillators into commuting functions $\code{A}\mapsto\code{Ac}$ thereby giving up the ordering. From now on, all expressions are understood to be in normal ordered form. It remains to perform the gauge group algebra, cf.~\linesref{lin:gauge-algebra}. Since we are only interested in the large $N$ limit and since the potential $V$ would not imply any interactions between $\algU(1)$ fields, we may as well work with the gauge group $\grU(N)$ instead of $\grSU(N)$. This reduces the computation time enormously. After all contractions are carried out, we discard the $\grU(1)$-part again in \lineref{lin:gauge-algebra:B}. Eventually we truncate to the $\algSU(2)$ subsector, i.e. keep only the oscillators $\Zd,\Z,\Wd,\W$.

The output of the program is the planar energy operator in the $\algSU(2)$ subsector spoiled by terms that are sub-leading in $\frac{1}{N}$ and a dominant vacuum contribution. The vacuum bubbles can be identified easily as they do not carry any oscillators, and we erase them by hand. The sub-leading terms are eliminated when we change to the language of permutation operators \eqref{eqn:permutations} as follows. We apply the computed energy operator (without vacuum contribution) to some long sample states and keep only the leading order in $N$. Then we apply an ansatz for the planar energy operator in terms of permutation operators to the same sample states and fit the coefficients of the ansatz such that the out-states for both applications match.

Due to a limited hard disk capacity (\Form{} writes large intermediate results to disk) the actual computation was performed in several parts, i.e. the program was run various times---each time for a different parts of the input \linesref{lin:enop}.

\singlespacing
\begin{lstlisting}[% 
%    caption={Normal ordering of energy operator in $\grSU(6)$ model},%
%    xleftmargin=-0.25\marginparwidth,%
    basicstyle=\footnotesize\tt,%
    numbers=left,%
    numberstyle=\tiny,%
    breaklines=true] 
symbol N,e,u,power,I;
autodeclare index m,n;
autodeclare index a=6,b=6;
function V,X,A,D,In;
cfunction Tr,Ac;
vector Zd,Z,Wd,W;
dimension N;

L [T2] = + D(-1,0)*V*In;                                     /*@\label{lin:enop:A}@*/
L [T4] = + D(-1,0)*V*D(+1,0)*V*In;
L [T6] = + D(-1,0)*V*D(+1,0)*V*D(+1,0)*V*In
         - D(-1,0)*V*D(+2,0)*V*D(-1,0)*V*In;
L [T8] = + D(-1,0)*V*D(+1,0)*V*D(+1,0)*V*D(+1,0)*V*In
         - D(-1,0)*V*D(+1,0)*V*D(+2,0)*V*D(-1,0)*V*In
         - D(-1,0)*V*D(+2,0)*V*D(+1,0)*V*D(-1,0)*V*In
         - D(-1,0)*V*D(+2,0)*V*D(-1,0)*V*D(+1,0)*V*In
         + D(-1,0)*V*D(+3,0)*V*D(-1,0)*V*D(-1,0)*V*In;       /*@\label{lin:enop:B}@*/

#message Substitute vertex      /*@\label{lin:substitute-vertex:A}@*/
#do x=1,4
id,once V = 1/2*( + X(a`x',m1`x')*X(a`x',m2`x')*X(b`x',m3`x')*X(b`x',m4`x')
                  - X(a`x',m1`x')*X(b`x',m2`x')*X(a`x',m3`x')*X(b`x',m4`x')
                )*Tr(m1`x',m2`x',m3`x',m4`x');
#enddo                          /*@\label{lin:substitute-vertex:B}@*/

#message Replace fields X by oscillators A
id X(a?,m?) = I*A(a,m,-1)-I*A(a,m,+1);
Id I^2 = -1;

#message Remove propagators and projectors     /*@\label{lin:eval-prop-proj:A}@*/
repeat;
* Commute propagators and projectors to the right
  id D(power?,e?)*A(a?,m?,u?) = A(a,m,u)*D(power,e-u/2);
* Evaluate propagators
  id D(1,0)*In = 0;
  id D(2,0)*In = 0;
  id D(3,0)*In = 0;
  id D(4,0)*In = 0;
  id D(power?{>=0},e?!{>=0,<=0})*In = In*(1/e)^power;
* Evaluate projectors
  id D(-1,0)*In = In;
  id D(-1,e?!{>=0,<=0})*In = 0;
endrepeat;                                     /*@\label{lin:eval-prop-proj:B}@*/

#message Normal ordering                       /*@\label{lin:nod:A}@*/
repeat;
  id A(a?,m?,-1)*A(b?,n?,1) = A(b,n,1)*A(a,m,-1) + d_(a,b)*d_(m,n);
endrepeat;
* give up the ordering, normal ordering implied
id A(?m) = Ac(?m);                             /*@\label{lin:is-nod}@*/
id In = 1;                                     /*@\label{lin:nod:B}@*/

#message U(N) Group algebra                    /*@\label{lin:gauge-algebra:A}@*/
repeat;
  id Tr(?i,m?,?j)*Tr(?k,m?,?l) = Tr(?i,?l,?k,?j);
  id Tr(?i,m?,?j,m?,?k)        = Tr(?j)*Tr(?k,?i);
  id Tr()                      = N;
endrepeat;
* Discard U(1) part now
id Tr(m?) = 0;                                 /*@\label{lin:gauge-algebra:B}@*/

#message Truncate to su(2)
id Ac(a?,m?,+1)*Ac(a?,n?,+1) = 0;
id Ac(a?,m?,-1)*Ac(a?,n?,-1) = 0;
id Ac(a?,m?,+1)*Ac(a?,n?,-1) = Zd(m)*Z(n) + Wd(m)*W(n);

cycl Tr;

print +s +f;

.end
\end{lstlisting}
\onehalfspacing

%%%%%%%%%%%%%%%%%%%%%%%%%%%%%%%%%%%%%%%%%%%%%%%%%%%%%%%%%%%%%%%%%%%%%%%%%%%%%%%%%%%%%%%%%%%%%%%%%%%%%%%%%%%%%%%%%%%%%%%%%%%%%%%%%%%%
%%%%%%%%%%%%%%%%%%%%%%%%%%%%%%%%%%%%%%%%%%%%%%%%%%%%%%%%%%%%%%%%%%%%%%%%%%%%%%%%%%%%%%%%%%%%%%%%%%%%%%%%%%%%%%%%%%%%%%%%%%%%%%%%%%%%
%%%%%%%%%%%%%%%%%%%%%%%%%%%%%%%%%%%%%%%%%%%%%%%%%%%%%%%%%%%%%%%%%%%%%%%%%%%%%%%%%%%%%%%%%%%%%%%%%%%%%%%%%%%%%%%%%%%%%%%%%%%%%%%%%%%%

\end{document}